\documentclass[preprint,superscriptaddress,showpacs]{article}
\usepackage{epsfig,cite}
\usepackage{amsmath,amssymb,graphicx}
\usepackage{ae}
\usepackage{aecompl}
\usepackage{psfrag}
\usepackage[margin=1.5cm,bottom=2.0cm,noheadfoot]{geometry}

\begin{document}

\title{Field-driven translocation of regular block-copolymers through a selective 
liquid-liquid interface}
\author{A. Corsi$^{1}$, A. Milchev$^{1,2}$, V.G. Rostiashvili$^1$, and 
T.A. Vilgis$^1$\\
$^1$Max - Planck - Institute for Polymer Research -
Ackermannweg 10,\\ 55128 Mainz, Germany\\
$^2$Institute for Physical Chemistry, Bulgarian Academy of
  Sciences,\\ 1113 Sofia, Bulgaria
}

\maketitle

\begin{abstract}
We propose a simple scaling theory describing the 
variation of the mean first passage time (MFPT) $\tau(N,M)$ of a regular 
block copolymer of chain length $N$ and block size $M$
which is dragged through a selective liquid-liquid interface by an external field $B$.
The theory predicts a non-Arrhenian $\tau$ vs. $B$ relationship which depends strongly
on the size of the blocks, $M$, and rather weakly on the total polymer length, $N$. The
overall behavior is strongly influenced by the degree of selectivity between the 
two solvents $\chi$.

The variation of $\tau(N,M)$ with $N$ and $M$ in the regimes of weak and strong 
selectivity of the interface is also studied by means of  computer simulations
using a dynamic Monte Carlo coarse-grained model. Good qualitative agreement with 
theoretical predictions is found. The MFPT distribution is found to be well
described by a $\Gamma$ -
distribution. Transition dynamics of ring- and telechelic polymers is also examined
and compared to that of the linear chains. 

The strong sensitivity of the ``capture'' time $\tau(N,M)$ with respect to 
block  length $M$ suggests a possible application as a new type of chromatography
designed to separate and purify complex mixtures with different block sizes of the
individual macromolecules.
 
\end{abstract}

\section{Introduction}

In a recent series of studies \cite{Corsi_JCP,Corsi_EPL,Corsi_macromol,Corsi_JPSB},
we have reported the results of comprehensive computer simulations which model
the behavior of hydrophobic - polar (HP) multiblock copolymers at a selective 
penetrable interface which divides two immiscible liquids, like water and oil, 
each of them being favored by one of the two types of monomers. As is well known
from experiment \cite{Clifton,Rother,Wang,Omarjee}, in the presence of selective interfaces
for which the energy gain for a monomer in the favored solvent is large, the hydrophobic 
and polar blocks of a copolymer chain try to stay on different sides of the boundary
between the two solvents, leading thus to a major reduction of the interfacial 
tension which has important technological applications, i.e. for compatilizers, 
thickeners or emulsifiers. Not surprisingly, during the last two decades the 
problem has gained a lot of attention also from theory 
\cite{Sommer1,Sommer2,Garel,Joanny,Denesyuk} as well as from computer experiment 
\cite{Balasz,Israels,Sommer3,Lyats,Chen}. While in earlier studies attention has been 
mostly focused on diblock copolymers \cite{Rother,Omarjee} due to their relatively 
simple structure, the scientific interest shifted later to {\em random} HP-copolymers 
at penetrable interfaces \cite{Sommer2, Garel,Joanny,Denesyuk,Chen}. 
In contrast, our investigations have focused mainly on the impact of block size $M$
on the static properties and localization kinetics of regular multiblock copolymers
at the phase boundary between the two immiscible solvents. We showed that these 
are well described  by a simple scaling theory \cite{Corsi_JCP,Corsi_macromol} 
in terms of  the total copolymer length $N$ (the number of repeating units 
in the chain), the block size $M$ (the number of consecutive monomers of 
the same kind), as well as the selectivity parameter $\chi$, that is, 
the energy gained by a monomer when moving into the more favorable solvent.

In the present work we extend our investigations to field-driven multiblock 
copolymers which cross the liquid-liquid interface and are temporarily
trapped for a typical time $\tau(N,M)$. Our simulation results reveal a rich
behavior of the drifting chains with qualitative differences in the two regimes
of weak and strong selectivity $\chi$. We demonstrate that a scaling theory
may be designed which captures the main qualitative features of the Mean First
Passage Time $\tau(N,M)$ in terms of chain length $N$ and block size $M$ and
provides insight into the complex behavior of the copolymer being dragged 
through the interface. We will show that the time $\tau(N,M)$ is very sensitive towards the block length $M$. This results can be viewed as a potential ground for the possible development of block - sensitive chramotography designed to separate  complex copolymer mixtures. It is worthy of note that the dynamic Monte Carlo simulation, used in our study, does not take into account the hydrodynamic interactions. The role of hydrodynamic effects might be very important and deserves future investigation.

In Section \ref{Theory} we introduce our model and derive the main analytical results
which describe the variation of translocation time with block size $M$ and force 
strength $B$ within the framework of Kramers approximation for the first passage
time. The main features of the Monte Carlo model which is used in our simulations
are briefly sketched in Section \ref{SIMUL} while in Section \ref{MC_res} we discuss
the most salient results of our computer experiments. We end the paper with
the main conclusions which the present investigation suggests.

\section{Theory of copolymer detachment from the interface}\label{Theory}

We now consider the  model of translocation dynamics where a growing fraction of the
chain
\begin{figure}[ht]
\begin{center}
\centerline{ \includegraphics[width=2.8in, angle=0]{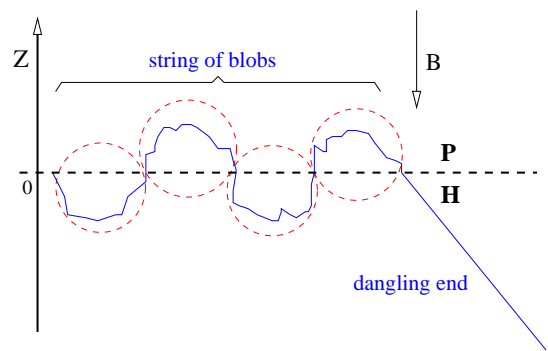} }
\caption{Schematic representation of the string of blobs subject to an external
force $ B $. 
In this scheme, the leading part of the chain (dangling end) is growing at the expense of the string of
blobs still captured at the interface. {\bf P} and {\bf H} denote the polar and hydrophobic semispaces respectively.}
\label{Tear}
\end{center}
\end{figure}
is detached from the interface by the external field $ B $ and dragged into the
underlying solvent as shown schematically in Figure \ref{Tear}. 
The physical picture which provides the basis for our detachment model will be
justified 
in Sec. \ref{MC_res} by the direct visualization within the Monte - Carlo simulation.
This model will be used here to describe the characteristic time $\tau$ it takes to
tear a 
string of blobs off the interface. To this end we calculate the free energy function. 

\subsection{Free energy of the detachment model}

We consider the free energy of the model, depicted in Figure \ref{Tear}, as a function
of the instant length of the chain, $m$, which is in the lower half (${\bf H}$) of the container (the dangling end length). 
The total free energy then contains the following terms:
\begin{equation}
{\cal F} (m) = \underbrace{F_{\rm sel}(m)  + F_{\rm el}(m)}_{\rm dangling \:\: end
\:\: free \:\: energy} + F_{\rm string}(m) \quad.
\label{Total}
\end{equation}
In this equation $ F_{\rm sel}(m) $ and $ F_{\rm el}(m) $ are the selectivity
interaction and 
the elastic deformation contributions of the leading end, whereas $ F_{\rm
string}(m) $ is 
associated with the energy of the remaining string of blobs.

The selectivity energy of the dangling end is related to the interface potential 
and the HP - copolymer binary sequence $ \sigma (s)$  ($\sigma(s)$ assumes the values $+1$ and $-1$ depending on the nature of the monomer, see ref.\cite{Corsi_JCP}.  
In the hydrophobic semispace ${\bf H}$ ($z < 0$) this free energy reads
\begin{equation}
F_{\rm sel}(m)  = \frac{\chi}{2} \int\limits_{N - m}^{N} d s \sigma (s)
\label{Selectivity}
\end{equation}
A simple  binary sequence for a multiblock copolymer with a block length of $ M $ 
can be constructed as $\sigma (s) = \sin \left( \frac{\pi}{M} \: s\right)$.
Substituting 
this in eq   \ref{Selectivity}  yields:
\begin{equation}
F_{\rm sel}(m)  = - \frac{\chi M}{\pi} \sin^2\left( \frac{\pi m}{2 M}\right)
\label{Selectivity2}
\end{equation}
where we have also taken into account that $N$ and $M$ have been chosen as multiples 
of $2$ (as is usual in MC - simulations).

\subsubsection{Free energy of elastic deformation}

Here we consider the deformation energy of the different parts of the chain based 
on scaling arguments. We assume the interface position to be at $z = 0$. The external 
field acts uniformly on each polymeric segment with a force $ B $ which points in the 
negative $ z $ direction.

Each instantaneous configuration can be thought of as an array of loops which spread 
on both sides of the interface. We denote them as positive $(z>0)$ and negative 
$(z<0)$ loops respectively. The two types of loops respond differently to the 
external field $B$. The positive loops are compressed whereas the negative loops are 
stretched with respect to their equilibrium shape for $B=0$ at the interface. 
Consider first the elastic free energy due to compression of a positive loop. 
Suppose a loop of a length $ n $ runs in the positive direction up to a height $ h $. 
Then the elastic free energy can be written as
\begin{eqnarray}
F^{(+)} (n, h)  \simeq T n \left( \frac{a}{h}\right)^{1/\nu} + \frac{2 B}{a} 
\int\limits_{0}^{h} z \: d z 
 \simeq  T n \left( \frac{a}{h}\right)^{1/\nu} + \frac{B h^2}{a}.
\label{Elastic_positive}
\end{eqnarray}
The first term in  the r.h.s. of eq  \ref{Elastic_positive} is due to confinement 
entropy whereas the second one is related to the external field energy, $ T $  
stands for the temperature and $ a $ denotes the Kuhn segments length.  
Optimizing eq  \ref{Elastic_positive} with respect to $ h $, one gets
$\frac{h_0}{a} = \left( \frac{n T}{a B}\right)^{\nu/(1 +2\nu) }$
which substituted back into eq  \ref{Elastic_positive} gives
\begin{eqnarray}
F^{(+)} (n) \simeq T \left( \frac{a B}{ T}\right)^{1/(1 + 2\nu)} 
n^{2\nu/(1 + 2\nu)}
\label{Elastic_positive_optim}
\end{eqnarray}

Let us now look at the stretching free energy associated with the deformation of the
a  negative loops. The elongation of a chain in a good solvent 
due to forces  $ f $  applied at  both ends \cite{Pincus}, 
for small extensions, $a n^\nu < h \ll a n$, reads
\begin{equation}
h \simeq  \left\{\begin{array}{l@{\quad,\quad}l}
a^2 n^{2\nu} f/T  &f \ll T/an^{\nu}\\
a n \left( a f/T\right)^{(1/\nu) - 1}  &f \gg T/a n^{\nu}
\end{array}\right.
\label{Stretching}
\end{equation}
Following F. Brochard - Wyart et al. \cite{Brochard1,Brochard2,Brochard3} (see also
the 
Monte - Carlo simulation in the wake of these results \cite{Milchev}), one
may describe the {\em non-uniform} stretching of tethered chains  in the differential
form,
$\frac{d h (s)}{d s} \simeq a \left( \frac{f a}{T}\right)^{(1/\nu) - 1}$ ,
where $ s $ is the running curvilinear coordinate of a polymeric segment 
and the field $ B $ acts on {\em each} monomer.  The total force is then $ f = s B $. 
The elongation of a string with $n$ monomers is then obtained from
$\frac{d h (s)}{d s} \simeq a \left( \frac{B s a}{T}\right)^{(1/\nu) - 1}$
by integration of $s$, $0\le s\le n $, and  yields
\begin{equation}
h \simeq a n^{1/\nu} \left( \frac{a B}{T}\right)^{(1/\nu) - 1} \quad.
\label{Integration}
\end{equation}
This result is valid for fairly large forces $ n B > T/a n^\nu $, 
or $ B > T/n^{\nu +1} $, which exceed the thermal agitation.
On the other hand, the Pincus' blob size can not be smaller than the Kuhn length, 
i.e. $ n B < T/a $, or $ B < T/a n $. As a result, the deformation law 
(\ref{Integration}) is valid for forces in the interval $ T/a n^{\nu +1} < B < T/a n $.
The corresponding free energy of stretching takes the form
\begin{eqnarray}
F^{(-)} (n) \simeq& \int\limits_{0}^{n} f \frac{d h}{d s} d s \simeq  
a \left( \frac{a B}{T}\right)^{(1/\nu) -1} B \int\limits_{0}^{n} s^{1/\nu} 
d s \simeq T \left( \frac{a B}{T}\right)^{1/\nu} n^{(1/\nu) + 1} \quad.
\label{Elastic_negative}
\end{eqnarray}

In the weak field regime, $ B < T/a n^{\nu + 1} $ the elongation law follows  from
the first line of eq  \ref{Stretching}, i.e.
$\frac{d h (s)}{d s} \simeq a \left( \frac{a B}{T}\right)s^{2\nu}$
and the resulting stretching free energy becomes
\begin{eqnarray}
F^{(-)} (n) \simeq& \int\limits_{0}^{n} f \frac{d h}{d s} d s  \simeq 
a \left( \frac{a B}{T}\right) B \int\limits_{0}^{n} s^{2\nu +1} d s 
\simeq  T \left( \frac{a B}{T}\right)^2 n^{2\nu + 2} \quad.
\label{Elastic_negative_weak}
\end{eqnarray}

Thus, the the elastic free energy in both weak, and relatively strong stretching 
regimes (see eqs  \ref{Elastic_negative_weak} and  \ref{Elastic_negative} ),  
is given by 
\begin{equation}
F^{(-)} (n) \simeq
\begin{cases}
T \left( \frac{a B}{T}\right)^2 n^{2\nu + 2}, &\text{if $ B < \frac{T}{a n^{\nu +1}}
$}\\
\\
T \left( \frac{a B}{T}\right)^{1/\nu}  n^{(1/\nu) + 1}, &\text{if $ \frac{T}{a
n^{\nu +1}} < B < \frac{T}{a n}  $}
\end{cases} \quad.
\label{Free_negativ_final}
\end{equation}

\subsubsection{Total free energy}

>From the previous section we have for the elastic free energy of the
leading part which includes $m$ monomers of the dragged chain,
according to eq  \ref{Free_negativ_final}, 
\begin{equation}
F_{\rm el} (m) \simeq - T \left( \frac{a B}{T}\right)^{1/\nu} \: m^{(1/\nu) +
1}\quad.
\label{Elastic_dangle}
\end{equation}
The negative sign here implies that  the chain moves downhill along the external field.

Consider now the free energy of the string of blobs, that is, of the rest of the 
copolymer which is still trapped at the interface. The length of such
a blob at the interface is \cite{Corsi_JCP}
\begin{equation}
g \simeq \chi^{-2/(1 - \nu)} M^{-(1 + \nu)/(1 - \nu)} \quad,
\label{Blob_size}
\end{equation}
whereby the blobs are placed alternatively in the $P$ - or $H$ - sides of the
interface. 
Each blob has an energy of the order of $ T $. The number of blobs is $(N - m)/g$ 
with $g$  - the number of (both $H$ and $P$) monomers in a blob.
Taking into account the expressions for the elastic energy
(\ref{Elastic_positive_optim}) 
and (\ref{Free_negativ_final}), the  free energy of the string of blobs is
proportional to
\begin{eqnarray}
F_{\rm string} (m) &\simeq& - T \left( \frac{N - m}{g} \right)  + \alpha_1  
T \left( \frac{N - m}{2 g} \right) \left( \frac{a B}{T}\right)^{\gamma} g^{1 
- \gamma} + \alpha_2 T \left( \frac{N - m}{2 g} \right) \left( 
\frac{a B}{T}\right)^{1/\nu} g^{(1/\nu) + 1}\nonumber\\
 &\simeq& - T (N - m) \left[ \frac{1}{g} - \frac{\alpha_1}{2} \left( \frac{a B}{T
g}\right)^\gamma - \frac{\alpha_2}{2} \left( \frac{a B g}{T }\right)^{1/\nu}
\right] \quad,
\label{String}
\end{eqnarray}
where $ \alpha_1 $ and $ \alpha_2 $ are some numeric factors and $ \gamma = 
1/(1 + 2\nu) $. The first term in eq  \ref{String} is associated with the 
energy of localization (therefore it is negative) whereas the two other terms are 
due to the elastic deformation of blobs.

In the strong localization limit the  positive and negative loops which constitute 
the string of blobs are of length $ M $. The localization energy for each 
loop is $ F_{\rm loc} \simeq - \chi M$ whereas the number of loops is $ (N - m)/M $. 
Thus, the free energy of the string of the loops reads
\begin{eqnarray}
F_{\rm string} (m) &\simeq& - \chi \left(N - m\right)  + \beta_1 T \left( 
\frac{N - m}{2 M} \right) \left( \frac{a B}{T}\right)^{\gamma} M^{1 - \gamma} 
+ \gamma_2 T \left( \frac{N - m}{2 M} \right) \left( \frac{a B}{T}\right)^{1/\nu} 
M^{(1/\nu) + 1}\nonumber\\
&\simeq& - T(N - m) \left[ \frac{\chi}{T} - \frac{\beta_1}{2} \left( \frac{a B}{T
M}\right)^\gamma - \frac{\beta_2}{2} \left( \frac{a B M}{T }\right)^{1/\nu} \right]
\quad,
\label{String_strong}
\end{eqnarray}
where $ \beta_1 $ and $ \beta_2 $ are constant factors.

With eqs \ref{Selectivity2},  \ref{Elastic_dangle},  \ref{String} and  \ref{String_strong}, the total  
free energy (eq  \ref{Total})  reads
\begin{eqnarray}
{\cal F} (m) \simeq &-& \frac{\chi M}{\pi} \sin^2\left( \frac{\pi m}{2 M}\right) 
-  T \left( \frac{a B}{T}\right)^{1/\nu} \: m^{(1/\nu) + 1}\nonumber\\
&-& T (N - m) \begin{cases}\left[ \frac{1}{g} - \frac{\alpha_1}{2} \left
( \frac{a B}{T g}\right)^\gamma - \frac{\alpha_2}{2} \left( \frac{a B g}{T }
\right)^{1/\nu} \right], &\text{for weak localization}\\
\\
 \left[ \frac{\chi}{T} - \frac{\beta_1}{2} \left( \frac{a B}{T M}\right)^\gamma 
- \frac{\beta_2}{2} \left( \frac{a B M}{T }\right)^{1/\nu} \right], 
&\text{for strong localization}
\end{cases}\quad.
\label{Total_free_energy}
\end{eqnarray}
The expression in square brackets of eq  \ref{Total_free_energy} (for the case of the weak localization 
limit) can be viewed as an inverse effective blob size $ 1/g $, i.e.
\begin{eqnarray}
\frac{1}{q} \equiv \frac{1}{g} - \frac{\alpha_1}{2} \left( \frac{a B}{T g}
\right)^\gamma - \frac{\alpha_2}{2} \left( \frac{a B g}{T }\right)^{1/\nu} \quad.
\label{Inv_blob}
\end{eqnarray}

The significant difference between the weak and strong localizations which is pointed out in eq \ref{Total_free_energy} will become apparent in the Sec. 2.2.4. Namely, there we will show that the characteristic translocation time for the weak localization regime grows monotone with the block length $ M $,  whereas in the strong localization case this dependence  passes through the maximum.

\subsection{Translocation time}

\subsubsection{Preliminaries}
It is well known that the  one-dimensional Smoluchovsky problem allows for an exact 
analytical solution for the mean first passage time \cite{Risken,Weiss,Kampen}. 
Let us assume that a stochastic process $ \xi(t) $ which describes the drift - 
diffusion of a chain in the external field $ U (x) $ extends between the boundaries $ x_1 $ or $x_2$,
 provided that $ \xi(0)= x_0 $. Then the expression for the mean passage time 
$ \tau (x_0) $ has the form
\begin{eqnarray}
\tau (x_0) = \int\limits_{x_0}^{x_2} d x' \: \frac{{\rm e}^{\beta U (x')}}{D (x')}
\; \int\limits_{x_1}^{x'} d y \: {\rm e}^{- \beta U (y)} \quad,
\label{Solution}
\end{eqnarray}
where $D (x)$ is the corresponding diffusion coefficient.
In view of the results of the previous section, we can map the problem at hand onto the one 
dimensional Smoluchovsky  problem where the corresponding  free energy ${\cal F} (m)$  
plays the role of an external potential. The question about the choice of the proper diffusion 
coefficient $D$ is more open. Considering the translocation  through a nanopore,
  Park and Sung\cite{Sung1,Sung2,Sung3} suggested that $D$ should be identified with the Rouse 
model diffusion coefficient which scales as $D = D_0/N$. On the other hand, 
Muthukumar \cite{Muthu} assumed that $D$ is not the diffusion coefficient  of the 
whole chain but rather the diffusion coefficient of the monomer which just passes 
the pore, and hence   a constant, independent of $N$ (see also an interesting 
discussion in ref. \cite{Milchev1}). We will consider here $D$ as a free parameter 
and will use a {\it correspondence principle} (see below) in order to fix it. 
We are interested in the average first - passage time when the length 
$ m $ changes within the interval $ 1 \le m \le N - 1$.  Using eq  \ref{Solution} 
this time  can be written in the form
\begin{eqnarray}
\tau (1 \rightarrow N - 1) = \frac{a^2}{D(N)} \int\limits_{1}^{N - 1} d m \: 
{\rm e}^{\beta {\cal F} (m)} \: \int\limits_{1}^{m} d n \: 
{\rm e}^{- \beta{\cal F} (n)} \quad.
\label{Trans_time_2}
\end{eqnarray}
We emphasize that this expression is basically valid under conditions when the chain 
is initially in equilibrium at the interface. The completely opposite case is that of a chain that is 
dragged through the interface while being in the steady state conformation typical of 
a drifting coil. The treatment of the latter case is more complicated and might  
not be simply based on the use of the same free energy unless the dragging motion is pretty
slow.

\subsubsection{Kramers' escape time}

One may simplify the double integral of eq  \ref{Trans_time_2} to treat it analytically.
To this end only two important terms in eq  \ref{Total_free_energy} can be left: 
the localization energy, $ - (N - m)/q $, and the elastic energy, $ - (a B/T)^{1/\nu} 
m^{(1/\nu)+1} $, the latter being responsible for the detachment. Then the resulting free energy 
function is shown in Figure {\ref{Resulting}.
\begin{figure}[ht]
\begin{center}
\centerline{ \includegraphics[width=3.0in, angle=0]{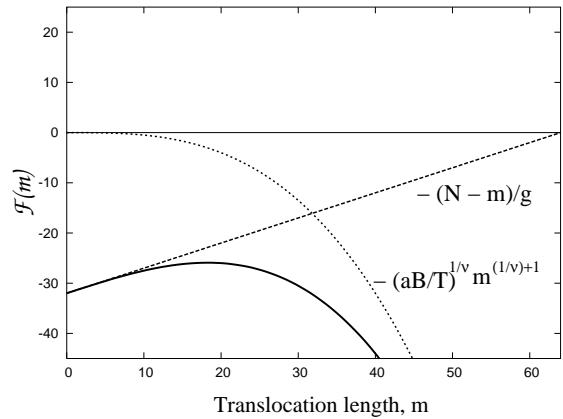} }
\caption{Schematic representation of the total free energy (solid line). The dashed 
line and the dotted line represent the localization energy and the elastic energy,
respectively.}
\label{Resulting}
\end{center}
\end{figure}

As one can see, the resulting barrier closely resembles that of the Kramers' escape problem 
\cite{Risken}. In order to roughly estimate the escape time and its dependence on 
$ B, N $ and $M$, we assume that the barrier determines the escape time. This means 
that the slowest part of the process is related to the barrier surmounting, 
whereas the rest of the translocation goes much faster. The free energy functional 
in this case has the form
\begin{eqnarray}
{\cal F} (m) \simeq - T \frac{N - m}{q} - T \left( \frac{a B}{T}\right)^{1/\nu} 
m^{(1/\nu)+1} \quad.
\label{Case}
\end{eqnarray}
which has a maximum at 
$m^* \simeq \frac{\gamma_1}{q^\nu }\left(\frac{T}{a B}\right)$
where the numeric coefficient $ \gamma_1 = 1/(1 + 1/\nu)^\nu $. The condition that 
$m^* $ cannot be larger than $ N $ leads to the restriction $ (a B/T) \geq 
\gamma_1/q^{\nu} N $. The corresponding free energy maximum value is given by
\begin{eqnarray}
{\cal F} (m^*) \simeq - \frac{N}{q} + \frac{\gamma_2}{q^{1+\nu} }
\left(\frac{T}{a B}\right) \simeq - \frac{N - m^*}{q} \quad,
\label{Max_value}
\end{eqnarray}
where $ \gamma_2 = 1/(\nu (1+1/\nu)^{1+\nu}) $.

The integral  over $ m $ in eq  \ref{Trans_time_2} gets its main contribution 
at $ m \simeq m^* $, so that we replace the factor 
$\exp\left\lbrace \beta {\cal F} (m)\right\rbrace $ by
$\exp\left\lbrace \beta {\cal F} (m^*) - \beta | {\cal F}''(m^*)| (m - m^*)^2/2 
\right \rbrace$, 
where $ \beta {\cal F}''(m^*) \simeq - (a B/T)^{1/\nu} 
(m^*)^{(1/\nu) - 1} $. Now, as $ m \simeq m^* $, 
the main contribution in the integral over $ n $ comes from the area $ n 
\simeq 0 $ 
(we have assumed for simplicity that the low integration limit  
equals zero). Then, $ \exp\left\lbrace \beta {\cal F} (n)\right\rbrace \simeq
\exp\left\lbrace - \beta {\cal F} (0)    - \beta {\cal F}' (0) n \right\rbrace $ 
where $  \beta {\cal F} (0) \simeq - N/q$ and $ \beta {\cal F}' (0) \simeq 1/q $. 
After taking all these expansions into account  the integrals in eq   
 \ref{Trans_time_2} can be easily calculated and the result has the following form
\begin{eqnarray}
\tau \simeq  q^{(3 - \nu)/2} \left( \frac{a^2}{D}\right) \left( \frac{T}{a B}
\right)^{1/2} \exp\left [\frac{1}{q^{1 + \nu}} \left( \frac{T}{a B}
\right) \right] \quad.
\label{Kramers}
\end{eqnarray}

The following comments are noteworthy in connection with this result. The exponential 
term has the form of an Arrhenius factor $ \exp(m^*/q) $ where $m^*/q$ is the 
barrier height (see the solid line in Figure\ref{Resulting}). The resulting $B$ - 
dependence of $\ln\tau$ appear as $ \ln\tau \sim T/(aB) $ which 
qualitatively agrees with the  MC - simulations findings (see below). At $ B 
\rightarrow 0 $ one has $ m^* \rightarrow N $ (it can be seen in
 Figure \ref{Resulting}) and the activation energy is simply equal to the 
localization free energy of the full chain $F_{\rm loc} \sim N/g  $ (see the 
corresponding result in ref.\cite{Corsi_JCP}) as it should be.

\subsubsection{Unperturbed drift}

As noted above, we will treat the coefficient $ D $ in eq  \ref{Kramers} as a free 
parameter which can be fixed by taking advantage of the {\it correspondence
principle}. 
In our case this principle states that for a sufficiently  large external force $ B
$ the 
activation barrier disappears and one observes a pure drift of the chain.
As one can see from eq  \ref{Total_free_energy}, this happens when $ 1/q \ll 1 $.
Then the interface crossing $ \tau_{\rm drag} $ is mainly defined by the balance 
of Stokes friction and external forces, i.e. $ \zeta_0 N v_{\rm c.m.} = N B $, 
where $ \zeta_0 $ is the Stokes friction coefficient and $ v_{\rm c.m.} $ is the 
velocity of the chain's center of mass, given by the formula $ v_{\rm c.m.} = 
B/ \zeta_0$. The effective thickness of the interface in this case is of the order 
of the blob length, i.e. $ \xi \simeq a g^{\nu} $ and the time $ \tau_{\rm drag} $ can 
be estimated as
$\tau_{\rm drag}  \simeq \frac{a g^{\nu}}{B}\: \zeta_0$
Then we can adopt $ \tau_{\rm drag} $ as a pre-exponent factor in the Kramers' 
escape time relation, i.e.
\begin{eqnarray}
\tau \simeq g^{\nu} \left( \frac{\zeta_0 a^2}{T}\right) \left( \frac{T}{a B}\right)
\exp\left\lbrace \frac{1}{q^{1 + \nu}} \left( \frac{T}{a B}\right) \right\rbrace
\quad.
\label{Kramers_1}
\end{eqnarray}

In the more general case we should deal with the complete full double integral 
of equation  \ref{Trans_time_2}. Once again, in this limit the second term in eq  \ref{Case} 
prevails and we have
\begin{eqnarray}
\tau  = \frac{a^2}{D} \int\limits_{0}^{N} d m \: \exp\left\lbrace -\left( 
\frac{a B}{T}\right) ^{1/\nu} m^{(1/\nu) + 1}\right\rbrace  \: \int\limits_{0}^{m} d n 
\: \exp\left\lbrace\left( \frac{a B}{T}\right) ^{1/\nu} n^{(1/\nu) + 1}\right\rbrace
\label{Double}
\end{eqnarray}
It is convenient to substitute the variables in both integrals with:
$ x = \left( a B/T \right) ^{1/\nu} m^{(1/\nu) + 1} $ and 
$ y = \left( a B/T \right) ^{1/\nu} n^{(1/\nu) + 1} $.
The resulting expression reads
\begin{eqnarray}
\tau  =\frac{a^2}{D} \left( \frac{T}{a B}\right)^{2/(1 +
\nu)}\int\limits_{0}^{\infty} d y \: \frac{{\rm e}^{-y}}{y^{1/(1+\nu)}}
\int\limits_{0}^{y} d x \: \frac{{\rm e}^{x}}{x^{1/(1+\nu)}} \quad.
\label{Double_1}
\end{eqnarray}
By setting the upper limit in the integral over $ y $ equal to infinity, we keep in mind 
that this integral converges as $ y \rightarrow \infty $ so that it does not depend 
on $ N $ at large $ N $. This convergence can be easily seen if we take into account 
the behavior of the integral over $ x $ for small and large $ y $ values, i.e.
\begin{eqnarray}
\int\limits_{0}^{y} d x \: \frac{{\rm e}^{x}}{x^{1/(1+\nu)}}\simeq
\begin{cases}
{\rm e}^{y}/ y^{1/(1+\nu)}, &\text{for $ y \gg 1$}\\
\\
y^{\nu/(1+\nu)} \quad , &\text{for $ y\ll 1  $}
\end{cases} \quad.
\label{Integral_x}
\end{eqnarray}
The correspondence to the {\it pure drift regime} leads us to the conclusion that the 
factor in front of the integral in eq  \ref{Double_1} should be equal to
$\tau_{\rm drag}$ (from $\tau_{\rm drag}  \simeq \frac{a g^{\nu}}{B}\: \zeta_0$). 
This enables us to fix the coefficient $D$ as
\begin{eqnarray}
D = \frac{1}{g^{\nu}} \left( \frac{T}{\zeta_0}\right) \: \left( \frac{T}{a B}
\right)^{(1 - \nu)/(1 + \nu)} \quad.
\label{D}
\end{eqnarray}

Finally, we need to point out that the equation for dragging time can be derived differently 
starting from the first passage time formula (\ref{Solution}) (see Appendix A).

\subsubsection{Numerical analysis}

For a first assessment of our theoretical treatment we present below some  numerical 
calculations based on the Kramers' escape time 
expression (\ref{Kramers_1}). First and foremost, we calculate the expression 
for the blob length $ g(\chi, M) $ which is of great importance in 
eq   \ref{Total_free_energy} for the total free energy. The number of monomers 
$g$ in the blob can be written as
$g(\chi, M) = N \left[ \frac{\chi_c(M, N)}{\chi}\right]^{2/(1 - \nu)}$ \quad,
\begin{table}[htb]
\begin{minipage}[b]{0.30\linewidth}
\begin{tabular}{|l|c|c|c|c|c|}
\hline
\multicolumn{6}{|c|}{Weak localization threshold $\chi_c$}\\
\hline
 & N=32 & N=64 & N=128 & N=256 & N=512\\
\hline
M=1 & 1.8003 & 1.5572 & 1.3467 & 1.1670 & 1.0083\\
\hline
M=2 & 1.5689 & 1.2902 & 1.0511 & 0.9591 & 0.9281\\
\hline
M=4 & 1.0157 & 0.9309 & 0.7379 & 0.6601 & 0.6526\\
\hline
M=8 & 0.7573 & 0.5851 & 0.4675 & 0.3405 & 0.2767\\
\hline
M=16 & & 0.2984 & 0.2900 & 0.1708 & 0.1720\\
\hline
M=32 & & & 0.2091 & 0.0829 & 0.0604\\
\hline
M=64 & & & & & 0.0524\\
\hline
\end{tabular}
\end{minipage}\hfill
\begin{minipage}[b]{0.46\linewidth}
\begin{tabular}{|l|c|c|c|c|c|}
\hline
\multicolumn{6}{|c|}{Strong localization threshold $\chi_{\infty}$}\\
\hline
 & N=32 & N=64 & N=128 & N=256 & N=512\\
\hline
M=1 & 3.7495 & 3.4189 & 2.9835 & 2.9022 & 3.2040\\
\hline
M=2 & 2.9507 & 2.8744 & 2.3954 & 2.1838 & 2.5536\\
\hline
M=4 & 2.0357 & 1.8855 & 1.6941 & 1.5616 & 1.6459\\
\hline
M=8 & 1.1484 & 1.1115 & 1.0688 & 1.0337 & 0.9187\\
\hline
M=16 & & 0.5593 & 0.5645 & 0.5358 & 0.5403\\
\hline
M=32 & & & 0.3603 & 0.2608 & 0.2941\\
\hline
M=64 & & & & & 0.1288\\
\hline
\end{tabular}
\end{minipage}
\caption{Weak and strong localization thresholds data}\label{Table22}
\end{table}
where the crossover selectivity $ \chi_c $ has been extensively discussed in ref.\cite{Corsi_JCP} 
and its scaling is $  \chi_c \sim M^{-(1+ \nu)/2} N^{-(1 - \nu)/2}$. The substitution 
of this $ \chi_c $ in $g$ leads back to eq  \ref{Blob_size}. The characteristic 
point where the blob length becomes of the order of the block length $M$ 
is   the strong localization threshold $ \chi_{\infty} $. Here we will use for  
$\chi_c $ and $ \chi_{\infty} $  the typical values obtained from Monte Carlo
simulations \cite{Corsi_JCP}. The corresponding data are given in Table \ref{Table22}. 
The subsequent calculations of the Kramers' escape time in the weak localization 
regime are done for a selectivity parameter $ \chi = 1.2 $.

We have used these data in the case $ N = 128 $  to make a 
3D - plot of $g$ which is shown in Figure \ref{3D_plot}. The contour lines in the
$\chi - M$ plane of Figure \ref{3D_plot} illustrate the narrow region of $M$-dependent 
selectivity strength where a crossover from weak to strong localization of the 
copolymer at the interface occurs. At the critical selectivity $\chi_c(M)$
the blobs reach the largest size $g\approx N$, which involves nearly all monomers
$N$, whereas with further increase of $\chi$ the blob size
\begin{figure}[ht]
\begin{center}
\centerline{\includegraphics[width=4.2in, angle=0]{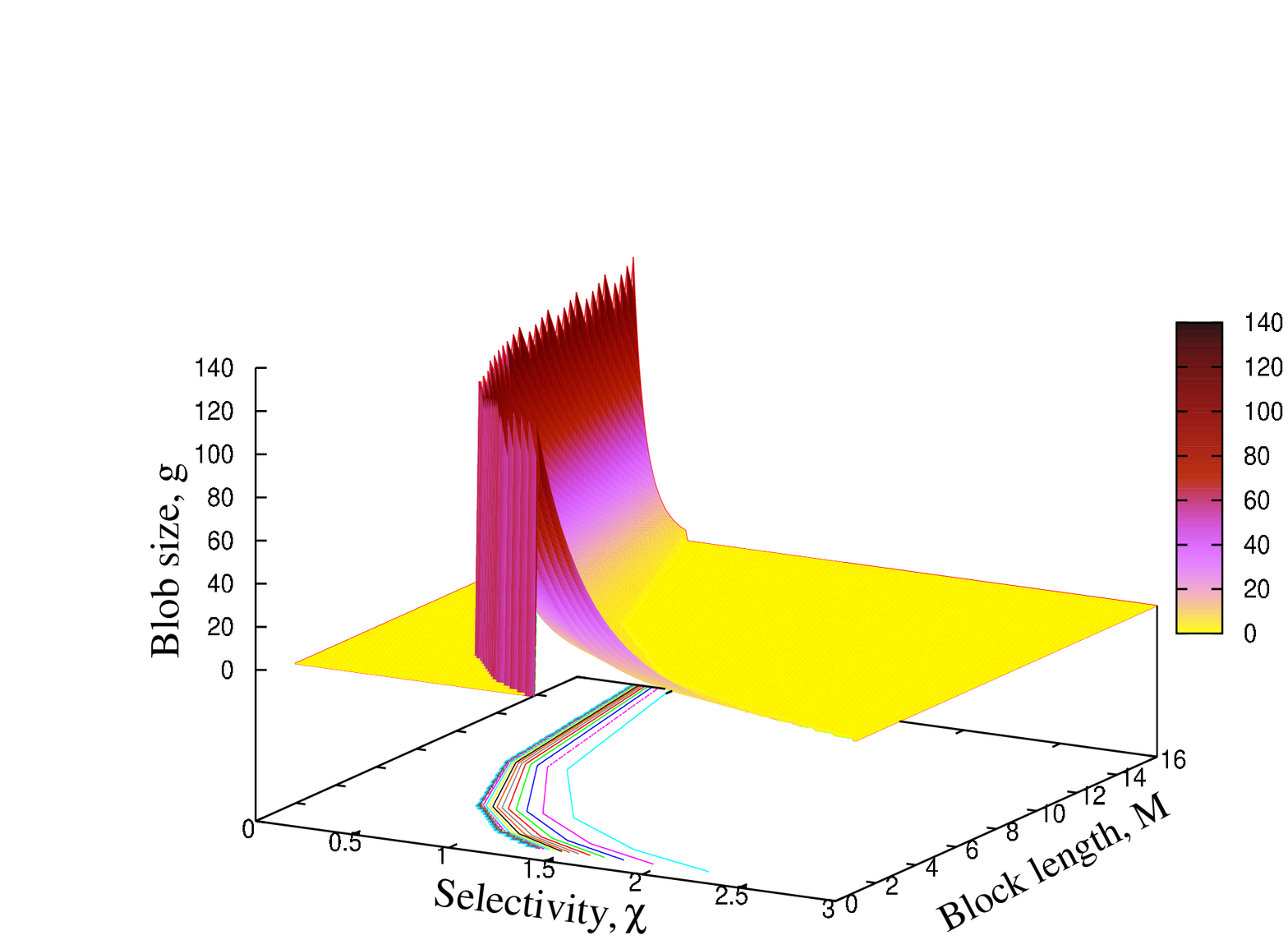} }
\caption{\label{3D_plot} 
Variation of the blob size $g$ with block length $M$ for various degrees of
interface selectivity $\chi $. The contour lines demark the region of
selectivity values in which a gradual transition from weak to  strong localization
takes place. At $\chi_c(M)$ (the first contour line on the left) one has
$g \approx N$ (here $N=128$)  while at $\chi_{\infty}$ (last contour line 
on the right) the blob coincides with the block size $M$.}
\end{center}
\end{figure}
rapidly decreases and at the onset of the strong localization, $\chi_{\infty}$, it reduces to the
size of an individual block $M$. As demonstrated below, within the 
narrow interval of weak localization, $\chi_c(M) \le \chi \le \chi_{\infty}(M)$, 
confined between the
outermost contour lines in Figure\ref{3D_plot} as well as in the vast region
of strong localization, $\chi \ge \chi_{\infty}$, both our theory and the 
simulations show a qualitatively different behavior of the translocation time 
$\tau$ with respect to $B$ and $M$ in the two regions.

The implementation of $ g(\chi, M) $ in the equation for Kramers' escape time 
(\ref{Kramers_1}) (where $ 1/q $ is given by eq   \ref{Inv_blob} ) leads to a
$ \tau $ vs. $ B $ relationship which is  given in Figure \ref{tau_B_plot}a.   
One can readily verify that this relationship is of non-Arrhenian type which
implies that the height of the activation barrier for crossing the interface 
varies with the variation of the dragging force $B$. Evidently, with growing
block size $M$ this non-Arrhenian behavior becomes progressively more pronounced.
We will see later that this is consistent with MC - results, and that the
non-Arrhenian variation of $\tau$ with $B$ becomes also progressively enhanced
as the chain length $N$ grows.

Figure\ref{tau_B_plot}b displays the variation of $\tau$ with $M$ for several values 
of the force $B$ in the regimes of weak and strong localization. One can see
that the $\tau$ vs. $M$ relationship differs qualitatively between the two cases.
In the weak localization regime one finds a very steep (by orders of magnitude) monotonous increase 
of $\tau$ with $M$, i.e. the escape time  is strongly selective with respect to 
the block size. In contrast, in the strong selectivity regime, $\chi= 4.0$, the
change of $\tau$ with $M$ is non-monotonous (it goes through a maximum) and is
not so strong. This is a result of the different responses of positive and 
negative loops to the external field $B$ as discussed in Section 2.1.2.

\begin{figure}[ht]
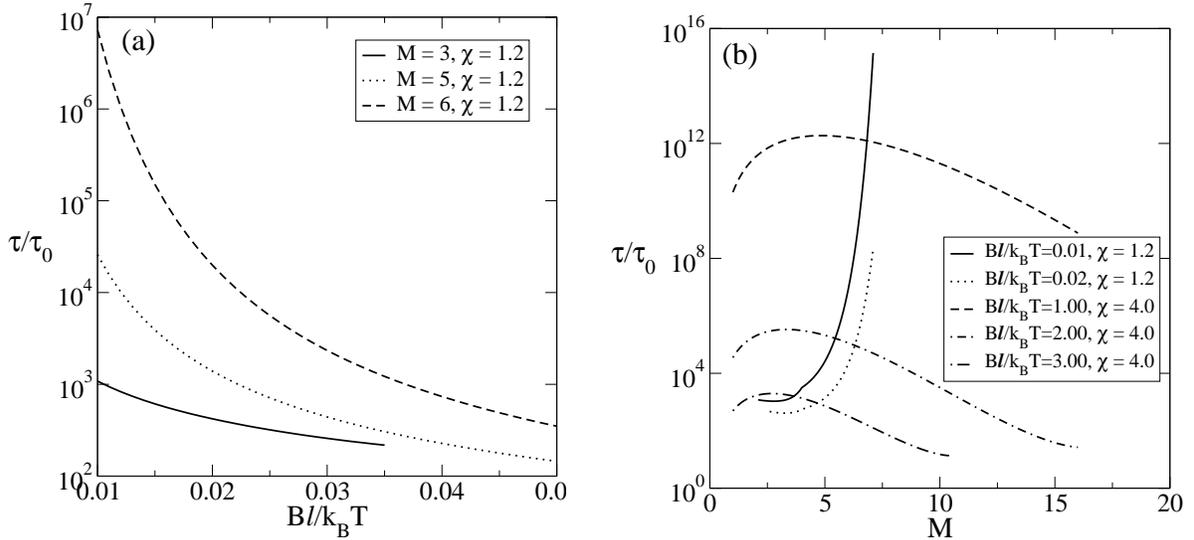

\begin{center}
\centerline{ 
\includegraphics[width=3.0in, angle=0]{tau_B_weak_andrea.eps}
\hspace*{5pt} 
\includegraphics[width=3.0in, angle=0]{tau_B_strong_andrea.eps}
}
\vskip 1.0cm
\caption{\label{tau_B_plot} 
(a) $\tau/\tau_0$ vs. dragging force $B$ for several block sizes $M$. 
(b) $\tau/\tau_0$ vs. $M$ in the regime of weak localization, $\chi= 1.2$,
and strong localization, $\chi= 4.0$, and several values of the external
field $B$.  }
\end{center}
\end{figure}

\section{Monte Carlo simulations}\label{MC}

\subsection{Simulation model}\label{SIMUL}

The off-lattice bead-spring model employed during our study has been previously 
used for simulations of polymers both in the bulk \cite{36,37} and near confining 
surfaces \cite{33,35,38,39,40,41}; thus, we describe here the salient features 
only. Each polymer chain contains $N$ effective monomers connected by anharmonic 
springs described by the finitely extendible nonlinear elastic (FENE) potential.

\begin{equation} \label{eq1}
U_{FENE} =-\frac{K}{2} R^2 \ln \big[1-\frac{(\ell - \ell_0)^2}{R^2} \big].
\end{equation}
Here $\ell$ is the length of an effective bond, which can vary in between 
$\ell_{min}< \ell <\ell_{max}$, with $\ell_{min}=0.4$, $\ell_{max}=1$ being the 
unit of length, and has the equilibrium value $\ell_0=0.7$, while $R=\ell_{max}
-\ell_0=\ell_0-\ell_{min}=0.3$, and the spring constant $K$ is taken as $K/k_BT
=40$. The nonbonded interactions between the effective monomers are described 
by the Morse potential
\begin{equation} \label{eq2}
U_M=\epsilon_M \{\exp[-2 \alpha(r-r_{min})]-2 \exp [-\alpha(r-r_{min})]\} \, ,
\end{equation}
where $r$ is the distance between the beads, and the parameters are chosen 
as $r_{min}=0.8$, $\epsilon_M=1$, and $\alpha=24$. Owing to the large value of the 
latter constant, $U_M(r)$ decays to zero very rapidly for $r>r_{min}$, and is 
completely negligible for distances larger than unity. This choice of parameters 
is useful from a computational point of view, since it allows the use of a very 
efficient link-cell algorithm \cite{42}.  From a physical point of view, these 
potentials eqs ~\ref{eq1}, ~\ref{eq2} make sense when one interprets the 
effective bonds as Kuhn segments, comprising a number of chemical monomers 
along the chain, and thus the length unit $\ell_{max}=1$ corresponds physically 
rather to 1 nm than to the length of a covalent $C-C$ bond (which would only be 
about $1.5 \text{\r{A}}$). Since in the present study we are concerned with the 
localization of a copolymer in good solvent conditions, in eq   \ref{eq2} we retain 
the repulsive branch of the Morse potential only by setting $U_M(r) = 0 \quad 
\mbox{for} \quad r > r_{min}$ and shifting $U_M(r)$ up by $\epsilon_M$.

The interface potential is taken simply as a step function with amplitude $\chi$,
\begin{eqnarray}
\label{eq3}
U_{int}(n,z)=
\begin{cases}-\sigma(n)\chi/2, &z > 0  \\
\sigma(n)\chi/2, &z \le 0
\end{cases}
\end{eqnarray}
where the interface plane is fixed at $z = 0$, and, as explained before, $\sigma(n) = \pm 1$ denotes 
a ``spin'' variable which distinguishes between P- and H- monomers. The energy 
gain of each chain segment is thus $-\chi$, provided it stays in its preferred 
solvent.

During each Monte Carlo update, a monomer is chosen at random and one attempts to 
displace it randomly by displacements $\Delta x, \: \Delta y,\: \Delta z$ chosen 
uniformly from the intervals $-0.5\le \Delta x, \: \Delta y,\: \Delta z\le 0.5$. 
The transition probability for such an attempted move is simply calculated from 
the total change $\Delta E$ of the potential energies defined in eqs \ref{eq1} 
- \ref{eq3} as $W=\exp(-\Delta E/k_BT)$. According to the standard Metropolis 
algorithm, the attempted move is accepted only if $W$ exceeds a random number 
uniformly distributed between zero and unity. Since our potentials are constructed 
such that chains cannot intersect themselves in the course of random displacement 
of beads, one does not need to check separately for entanglement restrictions. 

One should point out that the external field $B$, which imposes a
bias in the random hops performed by the chain monomers, should be taken in the
simulation sufficiently weak\cite{MiWiLa}, $B \Delta Z \ll k_BT$. Indeed, the field
${\bf B}$ whose only component is directed along the negative $Z$-axis enters as
an additional term in the Boltzmann factor, $\delta E = B\Delta Z$, which makes 
the energy a steadily decreasing function of $Z$. The Metropolis algorithm thus 
drives the polymer chain towards an unreached (and unreachable) minimum which  produces a uniform drift 
downwards. The average jump distance in a dilute system $\langle \Delta Z\rangle
\approx 1/4$, no matter how strong the applied field is chosen\cite{MiWiLa}. Thus, 
for large $B$ the probability of a jump {\em along} the field
quickly saturates at unity so that any further increase of $B$ would not really
lead to higher drift velocity of the polymer.

In the course of the simulation the starting configuration of the copolymer is 
relaxed for a period of time and then placed at the ceiling of our simulation
box before the field $B$ acting in vertical direction is switched on. The onset
of the translocation process is defined as the time when the first monomer of
the drifting chain touches the interface and the moment when the last monomer
leaves the upper half of the box then denotes the end of the process. Since the
time $\tau$ during which the chain remains in contact with the interface 
fluctuates considerably, for all chains of length $16 \le N \le 256$ we take
averages over $1000$ simulation runs.

\subsection{Discussion of Monte Carlo results}\label{MC_res}

In section \ref{Theory} we assumed a mechanism of copolymer detachment from the 
interface which have lead us to predict a non-Arrhenian variation of the time
$\tau$ with the imposed field $B$. We demonstrated that a selective liquid-liquid
interface can be very sensitive with respect to the block size $M$ of a
 multiblock copolymer chain. On the other hand, as shown in Appendix 
\ref{app}, the $N$ dependence of $\tau$ is rather weak and nearly vanishes in
the limit of long chains. In order to test the predictions of
the analytic theory and deepen our understanding of the overall picture of translocation through a 
liquid interface, we report in the present section the results of some extensive 
Monte Carlo simulation by means of the model described in Section \ref{MC}.

\begin{figure}[ht]
\begin{center}
\centerline{ 
\includegraphics[width=3.2in, angle=0]{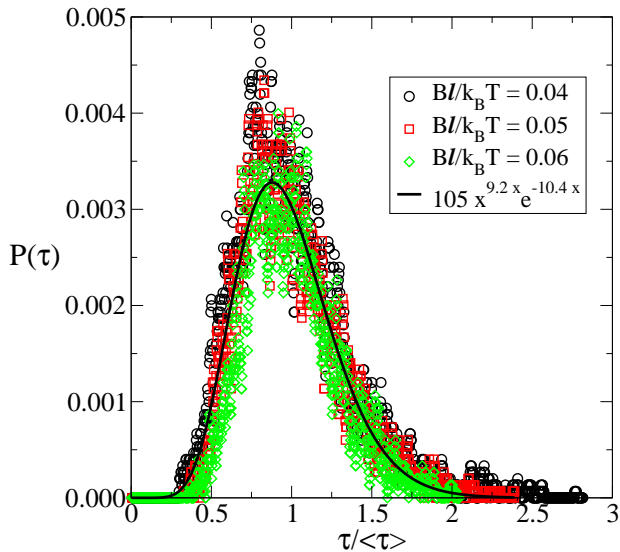} }
\caption{\label{PDF_fig} 
Master plot of the  distribution of capture times $P(\tau)$
vs. $\tau / \langle\tau\rangle $ for a copolymer chain with $N=128, \; M=4$ at
$\chi= 1.5$ for three different fields $B$, as indicated. The time intervals
have been scaled by the value of $\langle \tau \rangle$ appropriate for the respective value of the field 
$B$. A full line denotes the best fit of a $\Gamma$-distribution.
}
\end{center}
\end{figure}
In Figure \ref{PDF_fig} we show the distribution of capture times $P(\tau)$,
measured for a copolymer in the weak localization regime, for three 
different field strengths. Despite the considerable scatter in data which remains
after $1000$ measurements, evidently  these distributions are shown
to collapse on a single ``master'' curve when scaled by the mean capture time
$\langle \tau (B)\rangle$. The solid line in Figure \ref{PDF_fig} corresponds to a
$\Gamma$-distribution. This distribution arises naturally in processes
for which the waiting times between Poisson distributed events are relevant.

One of the main
results of our computer experiments is displayed in Figure \ref{tau_N} where
the variation of $\langle \tau \rangle$ with the bias $B$ is plotted for chains
of different length $N$ and fixed block size $M=4$ in semi-log coordinates.
A general feature, evident from Figure \ref{tau_N}a, is the well expressed
non-Arrhenian $\langle \tau \rangle$ vs. $B{\it l}/k_BT$ relationship which confirms
our expectations from Section \ref{Theory} - see Figure \ref{tau_B_plot}. 
>From Figure \ref{tau_N}a it becomes evident  that this non-Arrhenian type
relationship crosses over to Arrhenian one for shorter chain ($N=16\;,32$) where
insufficient length prevents the formation of (large) blobs of size $g$ as
expected at criticality $\chi_c= 1.5$. For sufficiently long copolymers, $N\ge 64$,
in contrast, the capture time $\tau$ depends only weakly on chain length $N$
and the $\tau$ vs. $B$ relationship remains qualitatively unchanged.  This
suggests that the activation barrier, which has to be overcome when a chain
is detached from the interface by the field, grows with decreasing drift velocity.
\begin{figure}[ht]
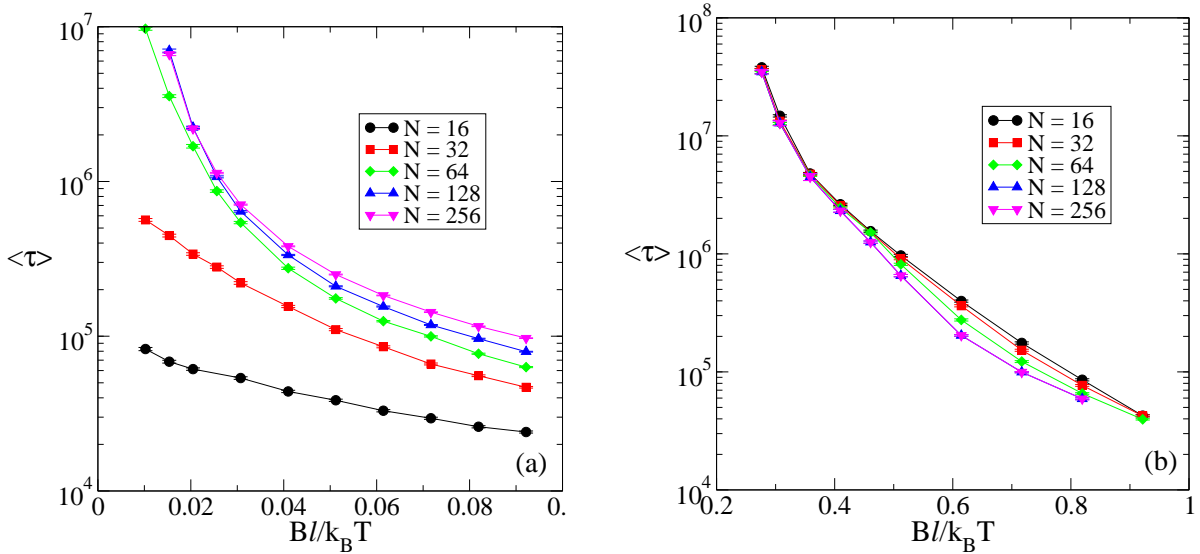

\begin{center}
\centerline{
\includegraphics[width=3.0in, angle=0]{tau_M4_weak_andrea.eps}
\hspace*{10pt}
\includegraphics[width=3.0in, angle=0]{tau_M4_strong_andrea.eps}
}
\caption{\label{tau_N}
(a) Variation of $\tau$ vs. dragging force $B$ for chain lengths $N=16\;,32\;,
64\;,128\;,$ and $256$ in the weak localization regime $\chi = 1.5$.
(b) The same as in (a) in the regime of strong localization, $\chi = 4.0$,
}
\end{center}
\end{figure}
One could imagine that in a strong field $B$ the copolymer moves so fast that
the translocation time is much shorter than the time it takes for the polymer to 
localize at the phase boundary  and attain its lowest free energy. For small 
values of $Bl/k_BT$, in contrast, the copolymer has sufficient time to attain
full equilibrium at the selective interface, and the corresponding potential well
in which the chain resides gets deeper. While the non-Arrhenian $\tau$ vs. 
$B{\it l}/k_BT$ relationship is also observed in the case of strongly selective
phase boundary, Figure \ref{tau_N}b, qualitatively the variation of $\tau$ with 
chain length $N$ looks quite different. While at both ends of the interval of
variation of $B$ the capture time practically does not depend on $N$, in the 
intermediate region, $0.4\le BL/k_BT\le 0.8$, the longer  chains spend less  time
at the interface than the shorter  ones! At present we lack a reliable explanation 
for this effect. One might assume, however, that in this interval of intermediate
$B$ intensities the process of copolymer detachment is governed by the probability
for the formation of a critical protuberance. Such a probability should then be 
proportional to the total chain length $N$. Considerations of this kind indicate, 
however, the necessity of a more detailed information about the underlying mechanism 
of copolymer detachment and its temporal evolution. Therefore, in Figure \ref{movie} we 
present a series of snapshots which capture the time history of detachment by
depicting the instantaneous positions of all repeat units of the chain at four
successive times after the onset of one particular translocation. One can 
clearly see in Figure \ref{movie} that after an initial latent period a protuberance
is formed at the free (dangling) end of the chain between monomers $120\le n\le 
128$ which from the start is immersed in its preferred solvent. After $20000$
MCS this end evolves into a progressively growing protuberance which eventually
tears off   the copolymer from the interface and drags it into the lower
half of the simulation box. Our observation show, however, that such protuberances 
form occasionally everywhere along the backbone of the chain and then die out 
with time or grow spontaneously like an avalanche. Thus the analogy with the
mechanism of detachment, envisaged in our theoretical treatment, Section \ref{Theory},
should be apparent. We note here that in order to check the role
which the dangling end of the chain plays in assisting the detachment process
we also made test runs with {\em ring} copolymers (see below) under otherwise
equivalent conditions and found no qualitatively different behavior when dangling
ends are eliminated.

\begin{figure}[ht]
\psfrag{n}{$\Theta_{\rm R}$}
\begin{center}
\includegraphics[width=2.0in, angle=270]{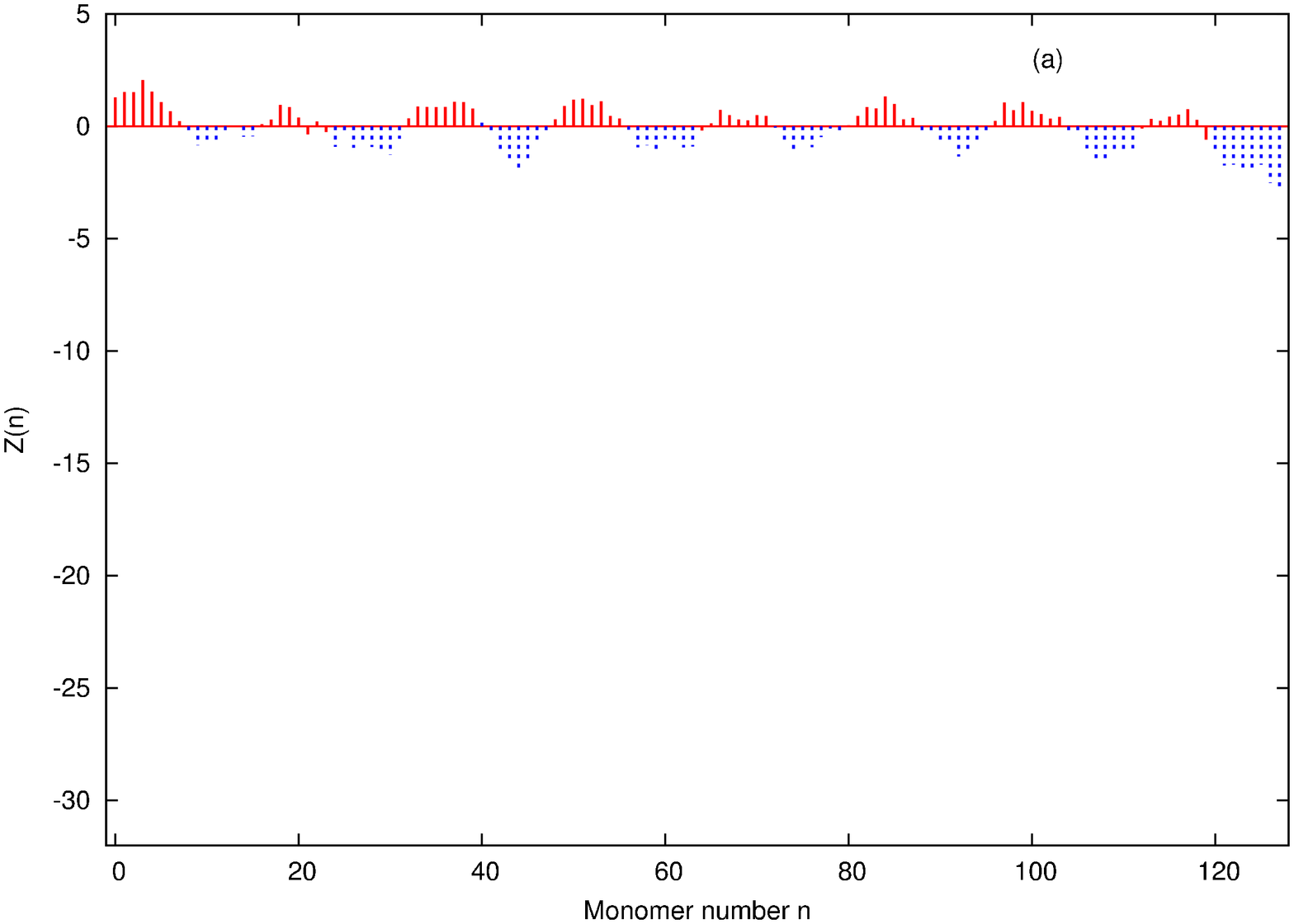}
\hspace*{10pt}
\includegraphics[width=2.0in, angle=270]{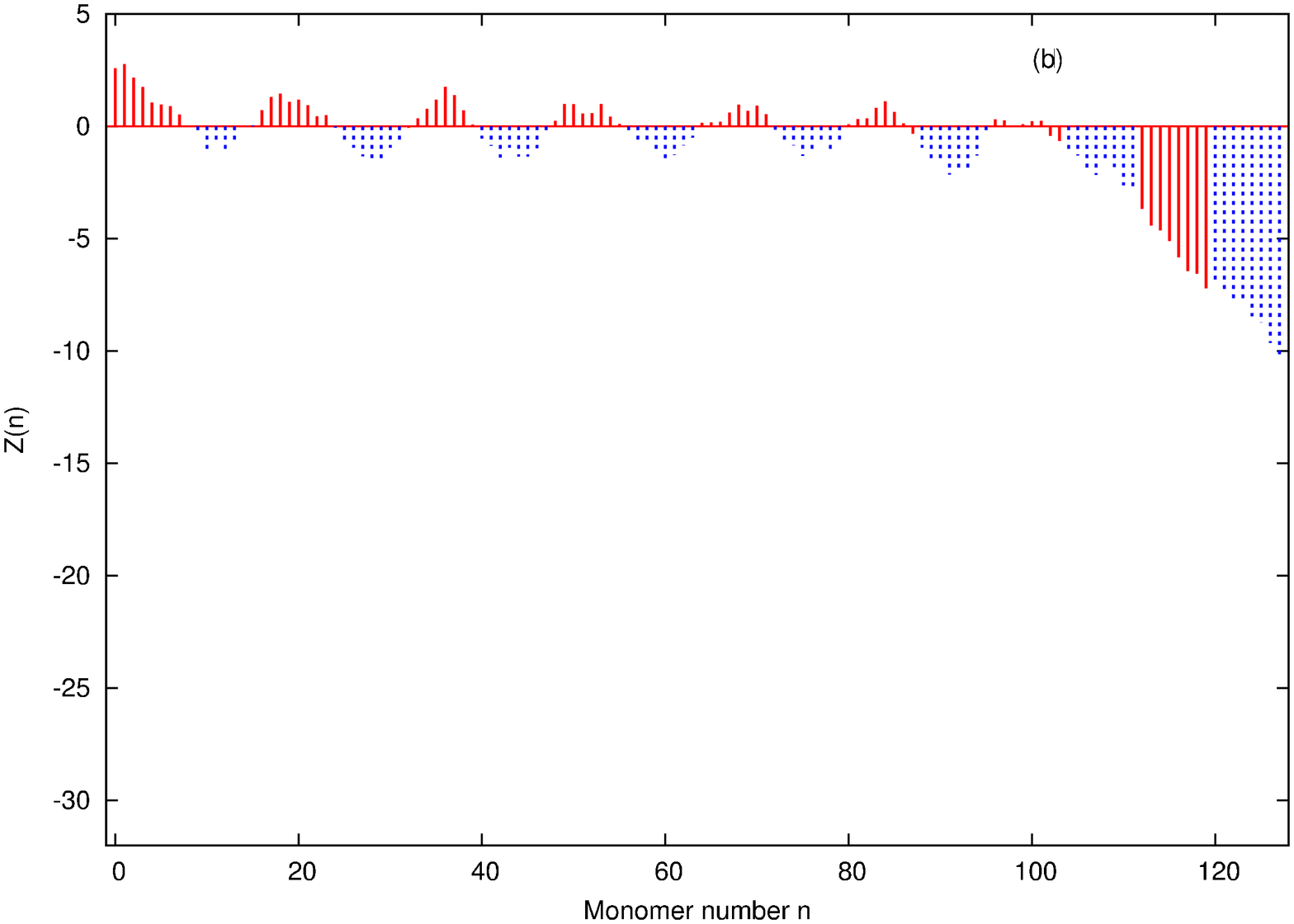} 
\\
\includegraphics[width=2.0in, angle=270]{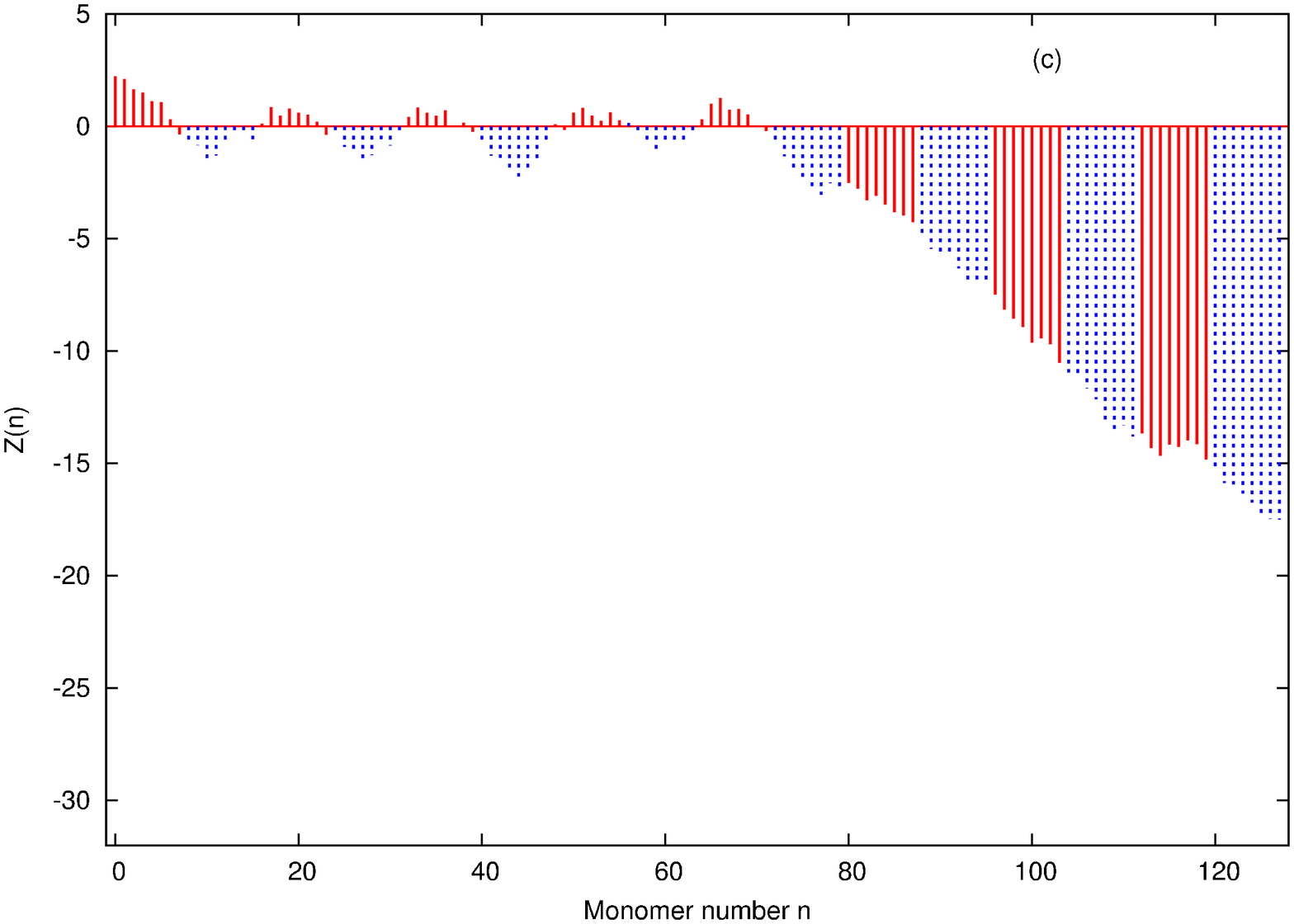}
\hspace*{10pt}
\includegraphics[width=2.0in, angle=270]{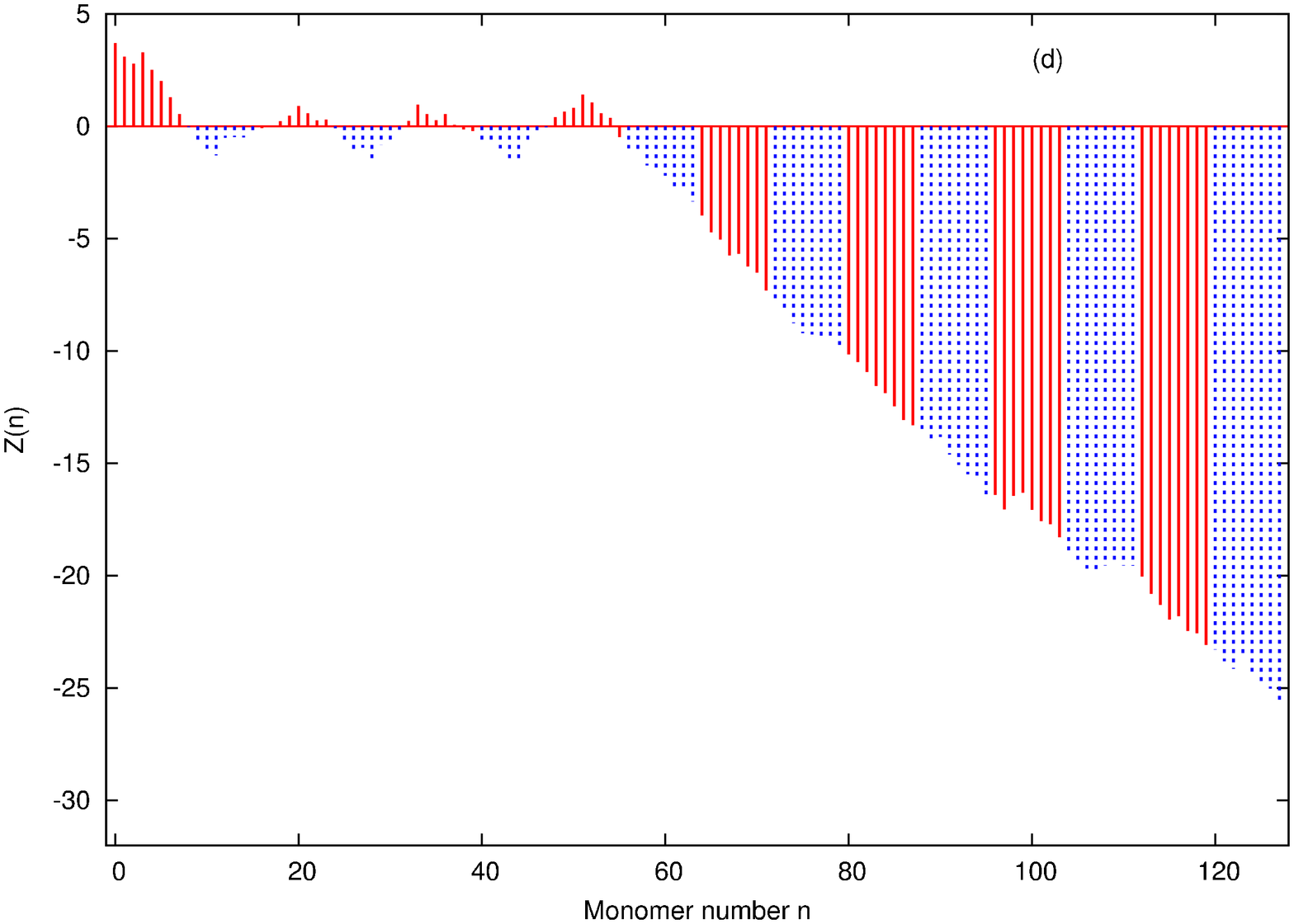} 
\vskip 1.0cm
\caption{\label{movie}
Snapshots showing the time evolution of the copolymer detachment from a liquid-liquid
interface for a chain with $N=128$ and $M=8$ (different blocks are represented by solid and dashed
lines) at $\chi = 1.5$. Displayed are the $Z$ -
coordinates of all $128$ monomers with respect to the position of the phase boundary
at $Z=0$. The times shown are at $325$ (a), $20325$ (b), $100325$ (c), and $200325$ 
MCS (d) after the start of the translocation process.}
\end{center}
\end{figure}
An important test for the theory, as developed in Section \ref{Theory}, is represented by
the analysis of the block
size dependence of the translocation time $\tau(M)$. In Figure \ref{tau_M_weak}a 
we demonstrate the extremely strong sensitivity of $\tau$ on $M$ for the case
of weak selectivity, $\chi= 1.5$, which leads to a steady increase in $\tau$
by more than a decade with doubling the block size. Figure \ref{tau_M_weak}a also
suggests that this sensitivity gets fully pronounced for sufficiently long
chains, $N\ge 64$, which we attribute to the formation of sufficiently large 
blobs unrestricted by the finite size of the copolymer. The variation of
the translocation time $\tau$ with block size $M$ for a fixed field strength
$B=0.3\;, 0.5\;, 0.7$ is shown in \ref{tau_M_weak}b where the steep (nearly
exponential) increase is evident. Qualitatively the observed behavior is well
in line with the one predicted by our theory and shown in Figure \ref{tau_B_plot}a.

For the case of strong selectivity, $\chi = 4.0$, again we find a qualitatively
different $\tau$ vs. $B$ behavior, see Figure \ref{tau_M_strong}, which agrees
quite well, however, with the theoretical predictions - Figure \ref{tau_B_plot}b.
For a given intensity of the dragging force one observes a non-monotonic change 
of $\tau$ with increasing block size $M$ with a local maximum around $M\approx 4$.
Thus one could conclude that our computer experiments provide a strong support
for the theoretical model, developed in Section \ref{Theory} on the ground of 
simple scaling considerations.
\begin{figure}[ht]
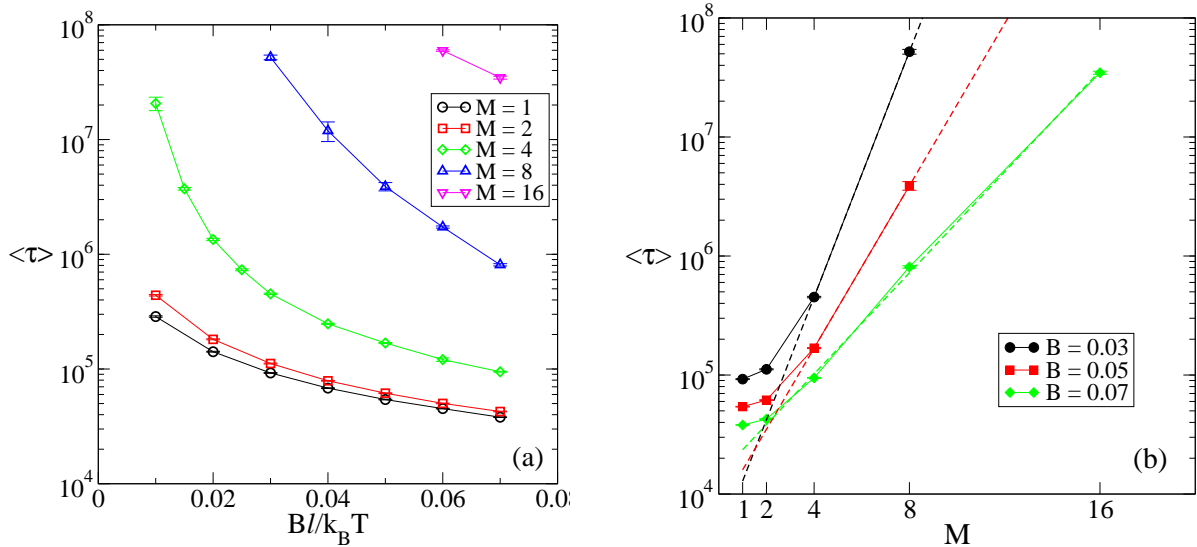

\begin{center}
\centerline{
\includegraphics[width=3.0in, angle=0]{weak_loc_vs_M_andrea_a.eps} 
\hspace*{10pt}
\includegraphics[width=3.0in, angle=0]{weak_loc_vs_M_andrea_b.eps}}
\caption{\label{tau_M_weak}
(a) Variation of the mean translocation time $\tau$ with field strength $B$ 
for copolymers of length $N=128$ and different block size $M$ in the weak
localization regime $\chi = 1.5$. (b) Variation of $\tau$ with block size $M$ 
for $B=0.03$, $B=0.05$ and $B=0.07$. An exponential fit is indicated by a dashed line.}
\end{center}
\end{figure} 

Eventually, before the end of this section we would like to point out several
additional results which are believed to underline the consistency of the
present combination of analytic and simulational results. An important question
concerns the impact of the starting conformation of the copolymer on the 
observed non-Arrhenian $\tau$ vs. $Bl/k_BT$ relationship. To this end in
Figure \ref{ring} we display a comparison of the measured $\tau(B)$ dependence
for chains which are localized and fully equilibrated at the liquid-liquid
interface {\em before} the drag force is switched on, and chains with identical
$N$ and $M$ which traverse the phase boundary ``on the fly''. As one may
readily verify from Figure \ref{ring}, for strong fields, $B\ge 0.03$, the chains
which start from equilibrated position {\em at the interface} need somewhat 
shorter time $\tau$ to cross it than those arriving from the upper half of the
simulation box. This appears reasonable since the former chains ``save'' the time 
interval between first touch of the interface whereby their center of mass 
still lags behind at $Z_{cm} \approx R_g$ above the separation line, and 
the moment when their center of mass arrives at the interface, $Z_{cm}\approx 0$.
For sufficiently small forces, however, one can see in Figure \ref{ring} that
both curves merge, that is, the time of residence at the interface is so long
that the small temporal advantage for a chain starting from the interface
is completely lost. Both in the case of ``equilibrated'' and ``on the fly'' chains,
however, qualitatively the $\tau$ vs. $Bl/k_BT$ relationship and its 
non-Arrhenian type remains unaltered which justifies the use of equilibrium 
free energy expressions for the copolymer in our theoretical treatment.

Another question which pertains to the specific mechanism of copolymer detachment
from the interface concerns the role of free (dangling) ends which are expected
to facilitate the process of detachment. A comparison between linear chains of
a given length, say $N=128$, and {\em ring} copolymers of the same length  might 
thus expose the role of free chain ends. From Figure \ref{ring} one can see
\begin{figure}[ht]
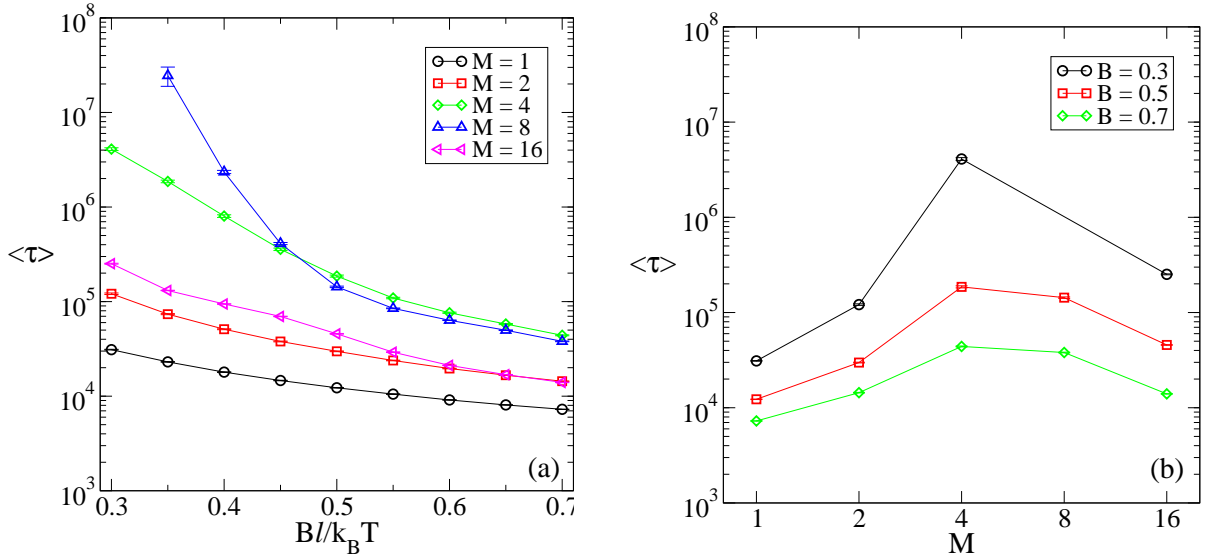

\begin{center}
\centerline{
\includegraphics[width=3.0in, angle=0]{strong_loc_vs_M_andrea_a.eps} 
\hspace*{12pt}
\includegraphics[width=3.0in, angle=0]{strong_loc_vs_M_andrea_b.eps}}
\caption{\label{tau_M_strong}
(a) Variation of the mean translocation time $\tau$ with field strength $B$
for copolymers of length $N=128$ and different block size $M$ in the strong
localization regime $\chi = 4.0$. (b) Variation of $\tau$ with block size $M$
for $B=0.3\;, 0.5\;, 0.7$. As expected, cf. Figure \ref{tau_B_plot}b, the
$\tau$ vs. $M$ relationship is for strong selectivity not a steady one.}
\end{center}
\end{figure}
that for strong forces, $B\ge 0.025$, the translocation time is about 50\% shorter 
in the case of ring whereas for $B\le 0.025$ it becomes gradually considerably
larger than for linear chains with dangling ends. Indeed, for strong fields 
when the objects do not have sufficient time to localize at the interface the 
ring polymer crosses faster the border line between the solvents because its 
size is smaller than that of the linear chain. In this regime the polymers move 
as almost compact objects and the presence of free ends does not affect the
the mechanism of interface crossing. In contrast, for weak forces the polymers
have sufficient time to localize and equilibrate on the interface while being
dragged through it. The effort to tear a dangling end off the interface would
then be only half of what is needed to create a protrusion in a ring copolymer
since the protrusion is hinged by two ends to the interface. Thus it should 
not be surprising that the detachment of ring copolymers take considerably 
longer than the that of the linear ones, as demonstrated by Figure \ref{ring}.

A special case which has not been in the focus of the present investigations
albeit it may serve as a check for our dynamic Monte Carlo simulation is 
shown in the inset of Figure \ref{ring}. There we present the variation of 
translocation time $\tau$ with chain length $N$ for a {\em telechelic} 
copolymer with fixed block size $M=4$ whereby a constant drag force $B=0.9$
is applied only to the head monomer of the chain. Evidently, one recovers
a quadratic increase of the translocation time with length $N$, $\tau 
\propto N^2$, which is qualitatively different from our data when the external 
field acts on all monomers of the chain. One can easily understand this 
observation, however, noting that for a telechelic chain the friction force
grows linearly with the number of beads $N$, meaning that the drift velocity
slows down as $N^{-1}$. In addition, the characteristic size of a stretched chain
should be proportional to $N$ too. The 
\begin{figure}[ht]
\begin{center}
\centerline{
\includegraphics[width=3.8in, angle=0]{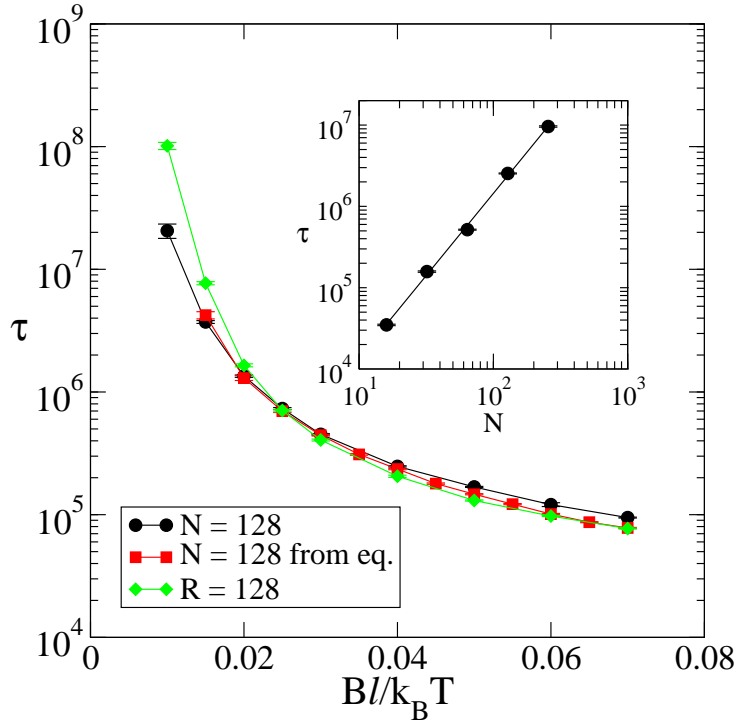} }
\caption{\label{ring}
Variation of $\tau$ with field strength $B$ for a chain with length $N=128$
and block size $M=4$. Circles denote a ring copolymer while squares mark linear
chains which start from equilibrated initial conformations localized at the
interface (full symbols), or enter the interface ``on the fly'' (empty symbols).
The inset shows the variation of $\tau$ with $N$ for a telechelic chain where
the field $B=0.9$ is applied only at the head monomer of the chain. The 
straight line denotes a power law, $\tau \propto N^\beta$, with an exponent
$\beta = 2.02 \pm 0.04$.
}
\end{center}
\end{figure}
combination of both factors should yield a quadratic growth of $\tau$ with
$N$, and this is indeed nicely confirmed by the simulation. This last 
result gives us an additional confidence in the consistency of our simulation
model and the observed results.

\section{Conclusions}

In the present work we have studied the transition of field-driven regular 
block copolymers through a selective liquid-liquid interface. We have developed 
a simple theory based on scaling considerations which appears capable of
catching the main features of the phenomenon, namely, the pronounced
non-Arrhenian dependence of the mean translocation time $\tau$ on field
intensity $B$, and its extreme sensitivity with respect to the copolymer 
architecture, that is, the block size $M$. It is based on the notion that the 
mechanism of copolymer detachment from the interface can be modeled as
an activation process whereby the copolymer is pulled out of a potential 
well, determined by the free energy of a localized equilibrium chain,
by a progressively growing protuberance. The height of the activation barrier
is shown to depend dramatically on the field strength $B$ - for large values
of $Bl/k_BT$ the barrier height disappears and the chain is dragged without
resistance through the selective phase boundary whereas for weak fields
the free energy of compressed and stretched blobs determines $\tau$. The
mean translocation time itself is tackled as a mean {\em first passage} time
using the Kramers approximation and thus predictions for qualitatively different
behavior of $\tau$ in the regimes of weak and strong selectivity are derived.

In order to check the analytic theory and provide a more comprehensive 
understanding of the process under investigation, we have reported  on extensive simulations
carried out by means of a dynamic Monte Carlo model which strongly support 
our findings. The simulations confirm the characteristic non-Arrhenian
type of $\tau$ vs. $Bl/k_BT$ relationship while revealing distinct differences
for the cases of weak and strong selectivity of the liquid-liquid interface.
One remarkable feature which is reproduced by the computer experiments is the 
strong sensitivity of the translocation time with respect to the block
size $M$ whereas the total chain length $N$ turns to have little effect on $\tau$.
This sensitivity suggests the interesting possibility to use selective liquid-liquid
interfaces as a new type of chromatographic tool whereby one can ``sieve'', that is, 
separate and analyze, complex mixtures of copolymers with respect to block size $M$. 
Recently a novel method to characterize individual blocks by means of Liquid 
Chromatography at the Critical Condition (LCCC)\cite{Jiang} has been proposed. 
In the present work we discuss another possibility of a new type of chromatography, 
based on the afore mentioned results.

One should note that in the rich behavior reported above we have not included results 
pertaining to {\em random} copolymers at penetrable selective interfaces where 
the range of sequence correlations plays a role similar to that of the block size 
$M$. Moreover, here we have totally ignored the possible effects of hydrodynamic 
interactions on the translocation kinetics. One might expect that the existence of a 
sharp interface, separating the two solvents, as well as the probable significant difference
in the solvents viscosities would generally suppress hydrodynamic effects across the 
phase boundary. Of course, the problem becomes more complex if the presence of a copolymer
at the interface changes the interface properties. These are important complications
which deserve a thought investigation. Nevertheless, we could expect that the main finding 
about the strong sensitivity of $ \tau $ with respect to $ M $ is unchanged. Eventually, 
this effect is evident from the localization free energy of the string of blobs (see eqs 
\ref{Blob_size}- \ref{String}), where the main term grows as $ F_{\rm string} \sim -T N M^{4} $, 
i.e. strongly depends on $ M $. Clearly, the verification of these 
predictions requires considerable computational efforts and therefore is on 
the agenda in our future work.

One should also keep in mind that all results have been derived and checked against
Monte Carlo simulations within the simplest model of an interface of zero thickness.
It is conceivable that by allowing for the presence of an intrinsic width of 
the interface as well as for a more realistic description involving capillary waves, 
etc., many additional facets of the general picture should become clearer. 
While these and other aspects imply the need of further investigations, i.~e., 
by means of single-chain laboratory experiments, we still believe that the present
study
constitutes a first step into a fascinating world which might offer broad perspectives
for applications and development.

\section{Acknowledgments}

AM acknowledges the support and hospitality of the Max-Planck Institute for
Polymer Research in Mainz during this study. This research has been supported by the
Sonderforschungsbereich (SFB 625).

\begin{appendix}
\section{Calculation of the pure drag time}\label{app}
In this Appendix we calculate the characteristic drag time for a polymer of  
length $N$ pulled by an external force $B$ along the $z$ - axis. The calculation 
is based on the first passage time formula (\ref{Solution})  which enables also 
to estimate $1/N$ - corrections. In this case the energy field in eq   \ref{Solution}  
has the form
\begin{eqnarray}
\beta U(z) = - \frac{N B}{T} \:z
\label{U}
\end{eqnarray}
The polymer chain center of mass starts from $ z = 0 $ at $t = 0$; we want to estimate the 
value of the time at which the center of mass of the chain passes 
$ z = L $. According to eq  \ref{Solution} the average time of the first passage reads
\begin{eqnarray}
\tau_{\rm drag} = \frac{\zeta_0 N}{T} \int\limits_{0}^{L} d z' \exp\left[ - 
\frac{N B}{T} \:z'\right] \int\limits_{0}^{z'} d z \: \exp\left[ \frac{N B}{T} \:
z\right]
\label{tau}
\end{eqnarray}
where we have used for the diffusion coefficient its Rouse expression $ D = T/
\zeta_0N $.
The straightforward calculation of the integrals in eq   \ref{tau} yields
\begin{eqnarray}
\tau_{\rm drag} = \frac{L \zeta_0}{B}\left\lbrace 1 - \frac{T}{N L B}\left[ 1 
- \exp\left( - \frac{N L B}{T} \right) \right] \right\rbrace
\label{tau_!}
\end{eqnarray}
One can see that for a large chain , i.e. at $ N \gg 1 $ and for a characteristic 
length of the order of the blob length, i.e. at $ L \simeq a g^{\nu} $ we go back to 
the simple expression $\tau_{\rm drag}  \simeq \frac{a g^{\nu}}{B}\: \zeta_0$. 
The finite chain size correction  can be expressed in terms of  
factors proportional to powers of  $ [1 - T/(B L N) ]$.

\end{appendix}

\end{document}